\documentstyle[twoside,fleqn,npb,epsfig]{article}
\newcommand{\AmS}{{\protect\the\textfont2
  A\kern-.1667em\lower.5ex\hbox{M}\kern-.125emS}}
\def\bb{{b\bar{b}}}
\def\cc{{c\bar{c}}}

\def\Abreu{DELPHI Coll., }
\def\etal{{\it et al.,}\ }

\def\b{$b$\ }
\def\bbg{$b\bar{b}g$\ }
\newcommand{\eps}{{\ifmmode \varepsilon        \else $\varepsilon$\fi}}
\newcommand{\thetac}{{\ifmmode \theta_C\else $\theta_C$\fi}}

\title{Jet and hadron production in e$^+$e$^-$ annihilations\\
{\small{To be published in the Proceedings of DIS 99 - Berlin 1999}}}

\author{Alessandro de Angelis\address{Dipartimento di Fisica dell'Universit\`a
Via delle Scienze 208, I-33100 Udine (Italia)}}%
\begin{document}

\begin{abstract}
The DELPHI detector at LEP is particularly suited for hadron identification 
and for the identification of $b$-initiated jets. This note presents recent results based on such characteristics.
\end{abstract}

\maketitle

\vspace*{-7cm}
\begin{tabular}{l r}
                       & \hspace{8cm} Dip. di Fisica dell'Universita' di Udine \\
                       & Preprint UDPHIR 99/01/AA \\
                       & 9 November, 1999 
\end{tabular}
\vspace*{7cm}

\section{Introduction}
This note covers three topics in the study of identified hadrons
and jet production, coming from recent analyses by the DELPHI experiment at LEP.
A description of DELPHI can be found in \cite{deldet}; its
performance is discussed in \cite{perfo}.

The selection of $q\bar{q}$ events at the Z peak (LEP 1) is easy: being the 
production of hadronic final states enhanced by three orders of magnitude
over the continuum, simple cuts against lepton-antilepton and two-photon 
events allow easily the definition 
of samples with contaminations at the
per mil level.

The situation is different at LEP 2 (above the Z). The 
hadronic cross-section is at the present energies of the order of 100
pb, and it is dominated by the radiative return to the Z peak.
The cross-section corresponding to effective c.m. energies above 0.85$\sqrt{s}$
is presently of the order of 20 pb, with a sizeable contamination
from WW events; thus, even with large luminosities, we
are left with a few hundreds of events per energy point (see the table below).
As a consequence, the cuts against contaminations cannot be too severe
(reaching purities of about 0.8, with efficiencies of about 0.8).
\begin{center}
\begin{tabular}{c|c|c}\hline
$\sqrt{s}$ (GeV) & Lumi/exp. & Events with \\ 
 & (pb$^{-1}$) & $\sqrt{s'}>0.85\sqrt{s}$\\ \hline
133 & 12 & 900\\
161 & 10 & 350\\
172 & 10 & 290\\
183 & 55 & 1,300\\
189 & 175 & 3,800\\ \hline
\end{tabular}
\end{center}

The system of hadron identification of DELPHI \cite{perfo} covers the
momentum region between 100 MeV/$c$ and 30 GeV/$c$, by the superposition of
the separations given by: the  
$dE/dX$ mainly in the region 
below 1 GeV/$c$; the liquid radiator of the Ring Imaging
CHerenkov Radiator (RICH) between 1 and 5 GeV/$c$; the gaseous
radiator of the RICH at high momenta. K$^{0}_{s}$ 
($\Lambda$) 
are reconstructed by invariant
mass distributions for opposite-sign particles coming from a secondary
vertex, 
with typical efficiency and purity of 30\% (25\%) and 95\% (90\%) 
respectively. 

The influence of the detector on the analysis is studied with
the full DELPHI simulation program, DELSIM~\cite{perfo}. The 
efficiency of identification of charged hadrons is computed directly from
the data, using samples of $\pi^\pm$ and protons coming from the
decay of K$^0_s$ and $\Lambda$.

\section{Identified hadrons in high energy $q\bar{q}$ events}

The way quarks and gluons transform into hadrons is not entirely
understood by present theories; the most satisfactory description is
given by Monte Carlo simulations. 
A different, analytical, approach (see e.g.
\cite{mlla}, \cite{khoze} and references therein) 
are QCD calculations using the so-called Modified Leading Logarithmic 
Approximation (MLLA) under the assumption
of Local Parton Hadron Duality (LPHD) \cite{mlla}. 
In this picture the particle yield is described
by a parton cascade, and the virtuality cut-off $Q_0$ is lowered 
to values of the order of 100 MeV, comparable to the hadron masses;
it is assumed that the results obtained for partons apply to hadrons as well. 

The momentum spectra of particles produced can be calculated
as a function of the variable $\xi_p$, where
$\xi_p = -\mathrm{log} (\frac{2p}{\sqrt{s}})$ 
($p$ is the momentum of the particle), and it depends on
three parameters: an effective scale 
parameter $ \Lambda_{eff} $, a momentum 
cut-off $Q_0$ in the evolution of the parton cascade
and an overall normalization factor $k$.  
The function
has the form of a ``humped-backed plateau'',  
  approximately  Gaussian in $\xi_p$~\cite{mlla}.

To check the validity of the MLLA+LPHD approach, 
it is interesting to study the
evolution with the centre of mass energy of the maximum,
$\xi^*$, of the $\xi_p$ distribution. In the 
framework of the MLLA+LPHD the dependence of $\xi^*$ on the centre
of mass energy can be expressed as:
\begin{equation}
\xi^* \simeq a + b \ln \sqrt{s}
\label{eq:xistar}
\end{equation}
where the slope $b$ depends just on the effective 
$\Lambda$ and not on $Q_0$: one can thus assume that $b$ is independent of the mass of the particle 
produced.

The evolution
of $\xi^*$ with the centre of mass energy for identified 
hadrons \cite{noi} was compared with lower energy data;
the data up to centre of mass energies of
91~$\mathrm{GeV}$ were taken from previous measurements~\cite{brummer}. 
The fit to expression~\ref{eq:xistar}
follows the data points rather well. 

The average multiplicity of the identified hadrons
was obtained from the integration of the $\xi_p$
distributions
inside a range varying according to the particle type and
energy; outside this range the fraction of particles
was extrapolated using the JETSET~7.4 \cite{lund} prediction. The
predictions from JETSET and HERWIG~\cite{HERWIG}, tuned at the Z, are
consistent with the observation.

\section{Charged particle multiplicity in $b\bar{b}$ events}

QCD predicts
that 
the difference in charge multiplicity between 
light quark and heavy quark initiated events in
$e^+e^-$ annihilations is energy independent; this is motivated by
mass effects on the hadronization (see \cite{khoze} for a review).
In a model in which the hadronization is independent of the mass,
this difference \cite{kisselev} decreases with the c.m.
energy, eventually becoming negative at LEP 2 energies.

Experimental tests at LEP 1 energies and below were not conclusive.
At LEP 2, the difference 
between the QCD prediction and the model ignoring mass effects 
is large, and the experimental measurement can firmly distinguish between the
two hypotheses. 
\begin{figure}[t]
\mbox{\epsfxsize8cm\epsffile{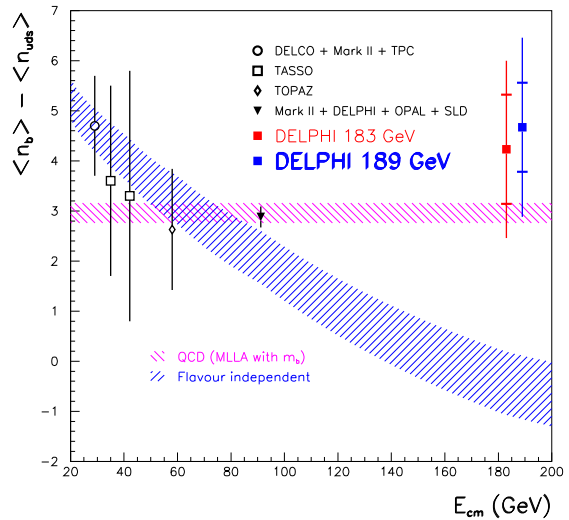}}
\vspace*{-0.95cm}
\caption[]{
The present measurement of $\delta_{bl}$ 
compared to previous measurements as a function of the centre-of-mass energy,
to the QCD prediction, and to the prediction from 
flavour-independent 
fragmentation.}\label{nice}
\vspace*{-0.3cm}
\end{figure}

Data collected by DELPHI 
at centre-of-mass energies around 183 GeV and 189 GeV were 
analysed~\cite{dn:bb}.
90\% $b$-enriched samples have then been obtained by 
tagging on the present of tracks with nonzero impact wrt the primary 
vertex~\cite{perfo}.
From such samples the average $\bb$ multiplicity has been measured by
unfolding via simulation for detector effects, selection criteria 
and initial state radiation. 
The difference $\delta_{bl}$ between the $\bb$ multiplicity 
and the multiplicity in generic light quark $l = u,d,s$ events 
was measured:
\begin{eqnarray*}
\delta_{bl}(183\,GeV) & = & 4.23 \pm 1.09 (stat) \pm 1.40 (syst) \, ;\\
\delta_{bl}(189\,GeV) & = & 4.69 \pm 0.89 (stat) \pm 1.55 (syst) \, .
\end{eqnarray*}
The result (Fig. \ref{nice})
is fully consistent with the
hypothesis of energy independence, and larger than
predicted by mass-independent fragmentation.
The systematic error 
will be further reduced by directly measuring the $\cc$ multiplicity.

\section{Identified hadrons in quark and gluon jets}

Jets originated from quarks  and  gluon 
are expected  to show  differences in their particle  multiplicity,
energy  spectrum, and angular  distributions, due to the different colour charges carried.
The LEP detectors can select gluon jets in \bbg events by tagging
the \b quarks.

2.2 MZ collected by DELPHI are
considered in the analysis~\cite{qgn}.
3-jet events are clustered using 
Durham with $y_{cut}=0.015$,  
optimizing the performance and still allowing a
reliable comparison with perturbative QCD.

For a detailed comparison 
with mininmum bias,  
it is necessary to obtain samples of quark  and gluon jets with
similar kinematics. To fulfill  this condition,
two different 3-jet event topologies have  been used:
{\bf{Y events}} (mirror symmetric), and
{\bf{Mercedes events}} (three-fold symmetric).

The number of Mercedes and Y 3-jet events was 
equal to 11,685 and 110,628 respectively.
After anti-tagging of heavy quark jets,
gluon jet purities of
$\sim 82\%$ for Y events and Mercedes events were achieved.
There are 24,449 Y events and 1,806 Mercedes
with identified gluon jets.

Three
different jet classes, namely normal mixture jets, $b$ tagged jets and
gluon tagged jets, with different compositions in terms of quark and gluon jets, were selected. From 
the comparison of these, the content in terms of protons, $K^\pm$,
$K^{*0}$ and $\phi$ was calculated as a function of the momentum of the hadron
for quark and gluon jets.
\begin{figure}
\begin{center}
\mbox{\epsfxsize6cm\epsffile{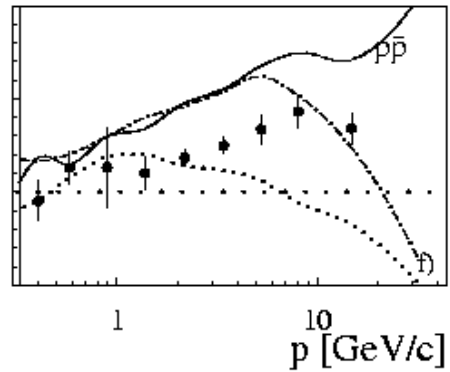}}
\end{center}
\vspace*{-1.0cm}
\caption[]{
Ratio of protons in gluon jets to quark jets, normalized to the ratio of
all charged particles, as a function of momentum.
}\label{iqg2}
\vspace*{-0.4cm}
\end{figure}

As for inclusive charged particles,
the production spectrum of identified hadrons is
softer in $g$ jets compared to $q$ jets, and the
multiplicity is larger. The ratio of the average multiplicity
in $g$ jets over $q$ jets
is consistent with the
ratio for charged particles, but for protons, where:
$$
 \frac{<n_{p}>^{(g)}/<n_{p}>^{(q)}}{<n_{ch.}>^{(g)}/<n_{ch.}>^{(q)}} = 1.205 \pm 0.041 \, .
$$
HERWIG underestimates both the $K$ and the p
production in $g$ jets; JETSET and ARIADNE
overestimate the proton production in $g$ jets.


\begin{thebibliography}{9}
\bibitem{deldet} \Abreu NIM {\bf A303} (91) 233
\bibitem{perfo}  \Abreu NIM {\bf A378} (96) 57
\bibitem{mlla} Y. Dokshitzer \etal
``Basics of Perturbative QCD'', Ed. Fronti\`eres, France 91 
\bibitem{khoze} V. Khoze and W. Ochs, Int.J.Mod.Phys. {\bf A12}
 (97) 2949
\bibitem{noi} DELPHI 99-21 MORIO CONF 220
\bibitem{brummer}  N. Brummer, Z.Phys. {\bf C66} (95) 367
\bibitem{lund} T. Sj\"{o}strand, Comp.Phys.Comm. {\bf 82} (94) 74
\bibitem{HERWIG} G. Marchesini \etal Comp.Phys.Comm. {\bf 67} (92) 465
\bibitem{kisselev} A. Kisselev, V. Petrov and O. Yuschenko, 
Z.Phys. {\bf C41} (88) 521
\bibitem{dn:bb} DELPHI 98-17 CONF 118 and
DELPHI prel. 
\bibitem{qgn} DELPHI 99-44 MORIO CONF 243
\end{thebibliography}
\end{document}